\documentclass[iop]{emulateapj}

\usepackage{graphicx,natbib,url,twoopt,amsmath}
\usepackage{hyperref}

\newcommand{\Hazel}{{\sc Hazel}}
\newcommand{\epsilonbold}{\mbox{\boldmath$\epsilon$}}

\shorttitle{Active region filaments might harbor weak magnetic fields}
\shortauthors{D\'iaz Baso, Mart\'{\i}nez Gonz\'alez \& Asensio Ramos}

\begin{document}

\title{Active region filaments might harbor weak magnetic fields}


\author{C. J. D\'{\i}az Baso, M. J. Mart\'{\i}nez Gonz\'alez and A. Asensio Ramos}
\affil{Instituto de Astrof\'isica de Canarias, C/V\'{\i}a L\'actea s/n, E-38205 La Laguna, Tenerife, Spain}
\affil{Departamento de Astrof\'{\i}­sica, Universidad de La Laguna, E-38206 La Laguna, Tenerife, Spain}

\begin{abstract} 
Recent spectropolarimetric observations of active region filaments
have revealed polarization profiles with signatures typical of the strong field Zeeman
regime. 
The conspicuous absence in those observations of scattering polarization and 
Hanle effect signatures was then pointed out by some authors. This was interpreted either 
as a signature of mixed "turbulent" field components or as a result of optical 
thickness. 
In this article, we present a natural scenario to explain
these Zeeman-only spectro-polarimetric observations of active region filaments. We 
propose a two-component model, one on top of the other. Both components have horizontal fields, the 
azimuth difference between them being close to 90 degrees. The component that lies lower in the 
atmosphere is permeated by a strong field of the order of 600 G, while the upper component has 
much weaker fields, of the order of 10 G. The ensuing scattering polarization signatures of the 
individual components have opposite signs, so that its combination along the line of sight reduces 
--and even can cancel out-- the Hanle signatures, giving rise to an apparent only-Zeeman profile. 
This model is also applicable to other chromospheric structures seen 
in absorption above active regions.
\end{abstract}



\section{Introduction} 

Solar filaments are dark thready structures seen on the disk as absorption 
in the core of some strong chromospheric lines (such as H$\alpha$ or \ion{Ca}{2} lines), 
other weak chromospheric lines such as the \ion{He}{1} multiplets at 10830\,\AA\ and 
5876\,\AA\ (D$_3$), and in the extreme ultraviolet continua. They are called prominences 
when they are seen as diffuse bright clouds at the limb, as they scatter light from the 
underlying disk. Broadly speaking, they can be segregated in Active Region (AR) and Quiescent (QS) 
filaments. The former lie above polarity inversion lines (PILs; the observationally defined 
line that delineates opposite polarity magnetic fields) of active regions. The latter lie above 
PILs in quiet Sun regions. QS filaments are very long structures 
that often live for weeks or even months, and that are suspended at heights up to 100 Mm 
\citep[][and references therein]{mackay2010}. AR filaments are formed in active regions, often in recurrent 
flaring areas, and are shorter in length and life time as compared to QS filaments. They are hardly seen 
as prominences at the limb because they probably lie lower in the atmosphere, at only a few Mm above the 
photosphere \citep[][and references therein]{mackay2010}.

From a physical point of view, solar filaments and prominences are cool chromospheric plasma overdensities 
embedded in the extremely hot and less dense corona. Magnetic fields play a fundamental role in the 
formation, support, and eruption of these structures. It is commonly agreed that these dense structures 
are formed in local dips of the magnetic field \citep[e.g.][]{KS1957,Aulanier2003A,Arturo2006}. 
However, the magnetism of solar filaments and prominences is very difficult to constrain observationally 
since it requires high precision spectro-polarimetric measurements and the interpretation of signals coming 
from the joint action of atomic polarization and the Hanle and Zeeman effects.

The \ion{He}{1} 10830\,\AA\ and D$_3$ multiplets are
very useful spectral lines to diagnose the dynamic and magnetic properties of
plasma structures at chromospheric temperatures. One of the advantages is that
the absorption of these lines is usually negligible in quiet regions, so that any 
absorption can be regarded as due to a levitating cloud at a certain height, 
which greatly simplifies the solution
of the radiative transfer equation for polarized radiation
\citep{leroy1977,sahal1977,TB2007}. In particular, these spectral lines have
been widely used for the study of solar prominences and filaments (e.g.
\citealt{Bommier1981,casini2003,merenda2006,orozco2014,marian2015}), the
reconstruction of magnetic fields in flux emerging regions
\citep{SOL2003,XU2010} and in chromospheric spicules (e.g.
\citealt[][]{ariste2005,centeno09,marian2012,orozco2015}).

Spectro-polarimetric observations of QS filaments and prominences revealed
magnetic fields that have strengths of the order of a few tens of G
\citep{TB2002,casini2003,merenda2006}, where the linear polarization of the \ion{He}{1} 10830\,\AA\ is dominated by the atomic level polarization due to the anisotropic radiation field. Yet, stronger fields 
are found for the same structures in active regions, in the range of 200-600 G 
\citep{Wiehr1991,SASSO2011}. Even larger field strengths (up to 800 G) were obtained 
when large, Zeeman-like linear polarization signatures were detected in AR filaments 
\citep{CK2009,CK2012,XU2012}. The work of \cite{CK2009} was the first to show
such linear polarization profiles of the \ion{He}{1} 10830\,\AA\ multiplet with the
surprising typical symmetric shape with three lobes of the transverse Zeeman effect. 

\cite{TB2007} and \cite{CASINI2009} soon noted that, even for 
such strong fields, AR filaments spectropolarimetric 
observations should show signatures of scattering polarization 
and the Hanle effect. 
The obvious solution of having a magnetic field inclined by the Van Vleck
angle (the angle that fulfills $\cos \theta_{VV}=1/\sqrt{3}$ and for which  the
contribution of scattering polarization vanishes, \citealt{landi_landolfi04})
in the whole filament looked clearly improbable \citep{CK2009}. 
\cite{TB2007} pointed out that when the optical thickness of the 
filament becomes larger than unity,
radiative transfer effects start to play a role. In this case, the radiation
field inside the filament becomes more horizontal (parallel to the solar
surface) and it can compensate to some extent the predominantly vertical
(along the radial direction) radiation field that is pumping the \ion{He}{1}
levels. Therefore, radiative transfer inside the filament produces a reduction
of the radiation field anisotropy, that leads to a strong reduction of the
scattering polarization and Hanle effect signals. 
\cite{CASINI2009} invoked the presence of a quasi-random
magnetic field coexisting with the organized magnetic field of a filament. The
isotropic component of the magnetic field strongly reduces the Hanle signal
produced by the deterministic field.

In this article, we propose a scenario based on a two-component model that naturally 
explains the predominantly Zeeman profiles observed in AR filaments. As a consequence of the model, we infer the presence of weak magnetic fields in AR filaments. The proposed scenario is extensive to other structures seen in absorption towards the solar disk in lines such as the
\ion{He}{1} multiplets (e.g., filaments, fibrils in emerging flux regions, \ldots).


\section{A model for absorption structures in the \ion{He}{1} multiplets}

To our knowledge, all inversions of spectro-polarimetric observations of absorption features in the 
\ion{He}{1} multiplets have been carried out using a single atmospheric component, as if the observed
signal was generated only in the cool plasma overdensities. Even so, these structures (QS and AR filaments, fibrils 
in emerging flux regions, \ldots) usually
have optical depths below or of the order
of unity\footnote{We measure the optical depth in the red component of the \ion{He}{1}
10830 \AA\ multiplet.} \citep{CK2009,CK2012,SASSO2011,XU2012}. Consequently, it is reasonable
to think that the emergent polarization signal can have some contribution
from underlying layers. This poses no problem in quiet regions because the
\ion{He}{1} multiplets yield almost no absorption. However, this turns out to be problematic
in active regions because, for instance, their chromospheres produce a significant absorption
in these multiplets.

We then propose a two-component approach to model the observed Stokes profiles 
of absorption features above active regions in the \ion{He}{1} multiplets. This
simplified model is made of two slabs with constant physical properties, one (slab 2) 
on top of the other (slab 1). Interestingly, 
this model allows us to naturally reproduce the Zeeman-only Stokes
parameters observed by \cite{CK2009} in AR filaments without any additional
mechanism to reduce the radiation field anisotropy. In such scenario, the emergent Stokes
parameters $\mathbf{I}=(I,Q,U,V)^\dag$ (with the $\dag$ symbol denoting
transpose) can be written as \citep{trujillo03}: 
\begin{align} \mathbf{I}_1 &=
e^{-\mathbf{K}^*_1 \tau_1} \mathbf{I}_\mathrm{sun} + \left( \mathbf{K}^*_1
\right)^{-1} \left( \mathbf{1} - e^{-\mathbf{K}^*_1 \tau_1} \right)
\mathbf{S}_1, \nonumber\\ \mathbf{I} &= e^{-\mathbf{K}^*_2 \tau_2}
\mathbf{I}_1 + \left( \mathbf{K}^*_2 \right)^{-1} \left( \mathbf{1} -
e^{-\mathbf{K}^*_2 \tau_2} \right) \mathbf{S}_2, \label{eq:twoComponents}
\end{align} 
where $\mathbf{I}_\mathrm{sun}$ is the Stokes vector that
illuminates the lower boundary of the slab (essentially the photospheric continuum), 
$\mathbf{K}^*=\mathbf{K}/\eta_I$
is the propagation matrix normalized to the absorption coefficient for Stokes
$I$, $\mathbf{S}=\epsilonbold / \eta_I$ is the  source function vector, with
$\epsilonbold=(\epsilon_I,\epsilon_Q,\epsilon_U,\epsilon_V)^\dag$ the
emissivity vector \citep[see][for more details]{landi_landolfi04,HAZEL2008} and $\mathbf{1}$ the
identity matrix. The previous expression considers two slabs with different
optical  depths ($\tau_1$ and $\tau_2$) and different magnetic fields.

\begin{figure*}[t]
\includegraphics[width=\textwidth]{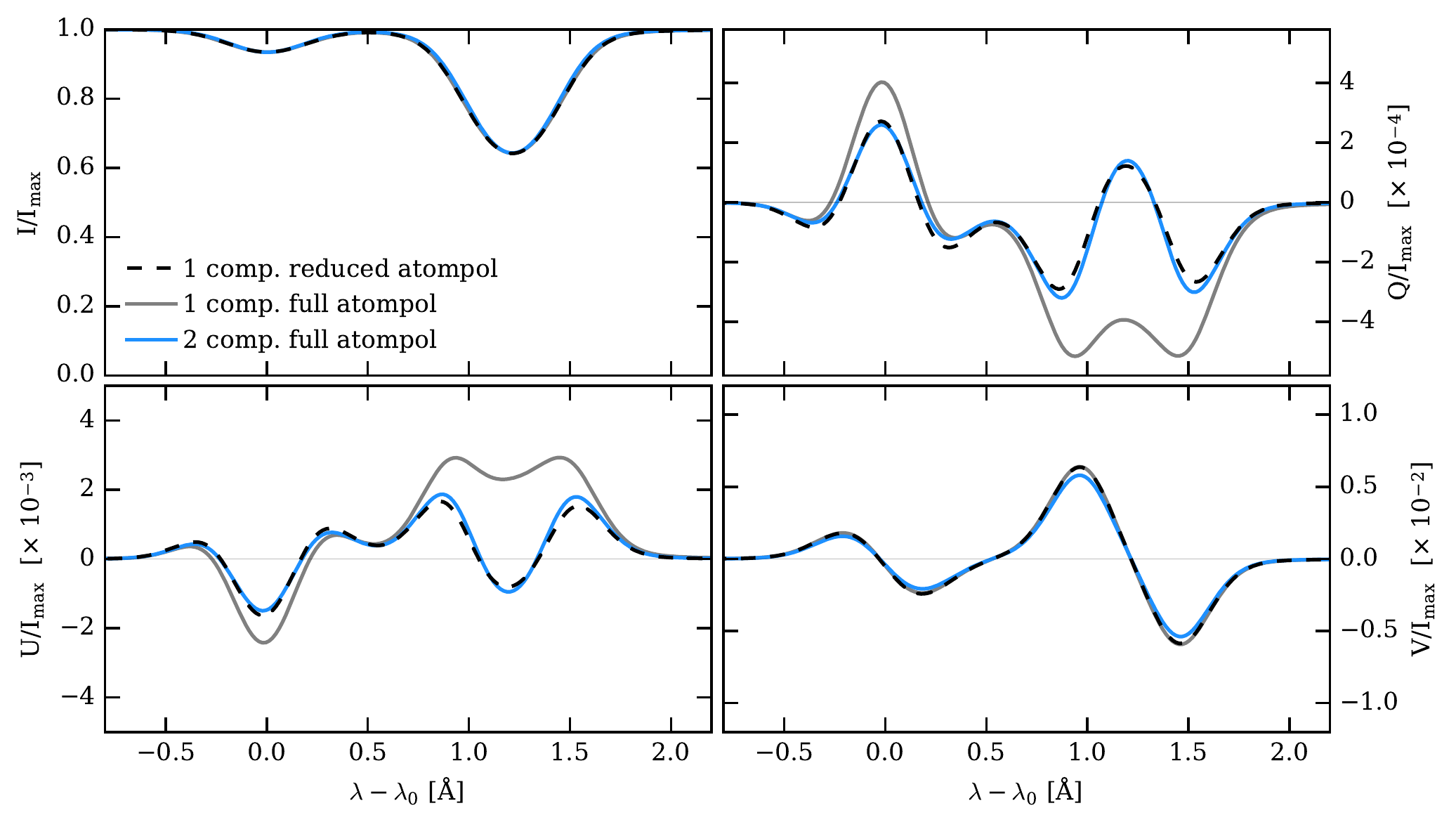}
\caption{Synthetic Stokes profiles with one component allowing atomic polarization but 
reducing the anisotropy by a factor of 0.2 (dashed black curve) that reproduces one of
the profiles inverted by \cite{CK2010}. The same profile but taking the full atomic
polarization into account (grey curve). The blue curve is the fit with two components.  
The reference wavelength $\lambda_{0}=10829.0991$ \AA\ 
is the central wavelength of the blue component of the \ion{He}{1} 10830 \AA\ multiplet. A color version of this figure is available online.
\label{fig:figure1}}
\end{figure*}

In some sense, our approach galvanizes from an idea proposed by
\cite{JUD2009}. This work suggested that the polarization signals observed in
the 10830 \AA\ multiplet in filamentary structures in emerging flux regions might
not mainly come from the highest part, but from the low-lying chromosphere of
the active region. \cite{JUD2009} argues that it is difficult to distinguish
the two possibilities because the Stokes profiles would look very similar.
Previous observations reinforce this idea because the photospheric
and chromospheric magnetic field maps obtained using inversion techniques 
look astonishingly similar. The chromospheric maps look slightly more fuzzy because 
of the reduction in the gas
pressure with height and the ensuing expansion, but no hint on any filamentary structure 
is detected \citep{SOL2003,XU2010,asensio2010}. We
go one step forward and stand on the idea suggested by \cite{asensio2010} to
propose that what we see is in fact a combination of the two atmospheres. We
do not need to distinguish the two options anymore because they are
simultaneously present in the Stokes profiles.

All calculations are done with \Hazel\footnote{The latest version is hosted at
\url{https://github.com/aasensio/hazel}.} \citep[HAnle and ZEeman Light,][]{HAZEL2008}, 
which is able to synthesize the \ion{He}{1} 10830\AA\ multiplet taking into account the presence
of atomic level polarization and the combined influence of the Zeeman and Hanle effects. 
The radiative transfer is carried out using a very simple slab model at a fixed height 
that assumes that all physical properties are constant within the slab. We use 
in this work the option of the code of two different slabs, so 
that the emergent Stokes parameters are computed using Eqs. \ref{eq:twoComponents}. Each slab 
is permeated with a different magnetic field vector but the pumping radiation in the two
slabs is the same. This simplification is probably not very realistic but this model suffices
to make our point (see Appendix for more details). 

To prove our idea, we synthesize the Stokes profiles using the parameters 
of one of the pixels along the filament displayed in Fig. 2 of \cite{CK2010}. The 
selected magnetic field has a strength of $B=550$ G, an inclination of $\theta_B=95^\circ$, 
and an azimuth of $\phi_B=49^\circ$, both angles defined in the local reference system 
(the quantization axis being the local vertical). Since the geometry of the field was provided in the observer's reference frame, we have transformed them to the local reference system taking into 
account that the heliocentric angle of their observations was $\theta=23^\circ$ ($\mu=\cos \theta=0.92$). 
Following \cite{CK2009}, our synthetic profile is obtained by artificially reducing
the anisotropy of the radiation field by a factor 0.2. 
We apply this correction to the radiation field anisotropy computed in \Hazel, which is 
obtained from the solar center-to-limb variation of the continuum at nearby wavelengths and correcting 
from the geometrical factor due to the height of the filament. The synthesized Stokes profiles are 
displayed in Fig. \ref{fig:figure1} with dashed black lines and are very similar to those 
published by \cite{CK2009}. We note that these profiles are almost indistinguishable 
from a scenario in which the presence of atomic level polarization is neglected. The grey lines show how the profile would look like using the very same physical properties
but fully taking into account atomic level polarization (i.e., using the radiation field
anisotropy computed by \Hazel). As already pointed out by \cite{TB2007}, clear signatures of 
atomic level polarization are seen even at fields above 1 kG, especially in the red component of the multiplet.

We carry out an inversion with \Hazel\ of the dashed black profiles 
shown in Fig. \ref{fig:figure1} with
a two-component model. The resulting Stokes parameters are displayed as solid blue
lines in Fig. \ref{fig:figure1}.  The inferred values for the low-lying
component are $B_1 = 650$ G, $\theta_1 = 92^\circ$, $\phi_1 = 48^\circ$, and $\tau_1 = 0.6$. The
upper component has $B_2 = 10$ G, $\theta_2 = 85^\circ$, $\phi_2 = 140^\circ$, and $\tau_2 = 0.3$. The fit is almost
indistinguishable from the profile with reduced anisotropy. Our inversion
indicates that the magnetic field of the low-lying component increases with
respect to the single-component inversion but its geometry is not modified
much because the Stokes $Q$, $U$ and $V$ signals are very strong. The magnetic
field in the upper slab turns out to be almost horizontal, with an azimuth that
is roughly perpendicular to that of the lower component and to 
the axis of the filament. We have carried
out extensive tests that indicate that the presence of a weak field in the
filament is a robust result. We stress the fact that the low-lying component has to
have a stronger magnetic field. Otherwise, the profiles cannot be fitted.  The 
magnetic configuration as well as the remaining slab properties were retrieved 
without imposing such a configuration as starting point of the inversion.

The profiles of the two-component model can be understood by noting
that the scattering polarization signatures generated in each individual component 
have opposite signs because the azimuths of the fields differ by $\sim 90^\circ$. 
The solution of the radiative transfer equation combines them reducing the 
Hanle contribution, with the possibility of even fully cancelling it out. 
We point out that the Hanle contribution to the line profile of both components
is similar even though the fields differ by more than an order of
magnitude because the \ion{He}{1} 10830 \AA\ is already in the Hanle saturation
regime for fields above $\sim 10-50$ G. In this regime, the Hanle signals are
insensitive to the magnetic field strength, so $B_2$ can be in this range.


\section{Stability of the solution} 

It is obvious that the proposed model presents a handicap
because a two-component model has a larger number of parameters and the
inversion becomes highly degenerate. For instance, \Hazel\ has some problems
converging to the same model when starting from different initial positions in
the parameter space. One of the most obvious degeneracies takes place on the
optical depth of each slab. We can only safely obtain the total optical depth
$\tau=\tau_1+\tau_2$ and the optical depth of each slab remains undetermined
provided their sum equals $\tau$. This affects the inference of the magnetic
field because it is sensitive to the specific value of the optical depth. The
only sensible way that we can think of to overcome this degenerate problem is
to use context information and take advantage of the whole observed map to
introduce constraints (e.g., use extrapolations from the photosphere to the
chromosphere).

In spite of the increase in the complexity of the model, we honestly
think that the model proposed in this paper is more physically 
realistic than any other single-component model
used in the past. Considering the increased degeneracy, it is very 
important to verify the consistency of the
model and check that the space of parameters that is compatible with the 
observations is not exponentially small. This would give us the idea that 
our fit is just a coincidence. To this end, we slightly perturb the original 
profile and inspect whether the inferred parameters do not vary much. 
The perturbed Stokes profiles are obtained by changing the magnetic field 
vector but keeping fixed the thermodynamic and dynamic properties 
of the single slab. We consider relative changes of 2, 5 and 10\% for the
three spherical components of the magnetic field vector. After that, we carry out 
an inversion with \Hazel\ and calculate the relative change in the 2-component 
model parameters with respect to the initial configuration.
The results are shown in Table \ref{tab:tabla1}. The table indicates the relative 
change in the magnetic fields of the upper and lower components when the magnetic 
field vector in the single-component is modified with a certain relative change. 
All horizontal lines refer to changes below 0.5\%, that is compatible with no 
change at all. We note that all changes are roughly of the same order than 
the modification in the original profile. Obviously, larger modifications of 
the profile lead to larger modifications of the 2-component model parameters. 
The important point of this table is that our proposed model is very robust to 
changes in the profile. Of special relevance is the fact that the azimuth difference 
between the two components is always $\sim 90^\circ$  for each perturbation. 
Converging to the solution with a two-component 
model is tougher and slower. For that reason, some values of the Table 
\ref{tab:tabla1} are obtained by setting the initial values to the original ones.
\begin{table}[t]
\centering
\caption{Relative change in the model parameters of the 2-component model when the 
magnetic field vector of the single-component model is modified. The horizontal 
lines indicate relative changes below 0.5\%, compatible with
no change at all.\label{tab:tabla1}}
\begin{tabular}{c|ccc|ccc|ccc}
\hline\hline
{$\Delta[\%]$} & \multicolumn{3}{c|}{2\%} & \multicolumn{3}{c|}{5\%} & \multicolumn{3}{c}{10\%}\\
\hline
                     & $B$    & $\theta$  & $\phi$ & $B$   & $\theta$   & $\phi$& $B$  & $\theta$ & $\phi$\\
\hline
$B_1$                & 1      & -     & -     & 4     & -     & -    & 10    & -     & -      \\ 
$\theta_1$           & -      & 2     & -     & -     & 6     & -    & -     & 12    & 1      \\ 
$\phi_1$             & -      & -     & 1     & -     & -     & 4    & -     & -     & 8      \\ \hline
$B_2$                & 2      & -     & -     & 5     & -     & -    & 12    & -     & -      \\ 
$\theta_2$           & -      & 2     & -     & -     & 4     & 5    & -     & 14    & 7      \\ 
$\phi_2$             & -      & -     & 1     & -     & -     & 3    & -     & 2     & 6       \\\hline
\end{tabular}
\end{table}

\section{Discussion and Conclusions}

We have presented a natural way to explain the absence of atomic level polarization signatures in
some observed AR filaments. This absence was interpreted in the past as an indication that the
magnetic field in AR filaments was very large, well above 500 G. Additionally, an extra mechanism
had to be invoked to destroy the, otherwise present for these fields, signatures
of atomic polarization and the Hanle effect. 

We propose a two-component model, one on top of the other\footnote{ Two atmospheres combined
with a filling factor could also be an option but this physical configuration still has to be justified, like other potential scenarios as the one presented by \cite{TB2010}.}. 
At certain configuration of the magnetic field in the two components, 
the scattering polarization and the Hanle effect signals can cancel out. Therefore,
we do not need any 
additional mechanism to reduce the anisotropy. To show this idea,
we inverted a synthetic profile computed from the results of \cite{CK2010}. 
We infer a weak magnetic field (around tens of G) in the top component, and 
a strong hG field in the bottom one. Both are horizontal but have an azimuth 
difference of around 90$^\circ$. 
A filament harboring weak fields suspended above an active 
chromosphere is then a plausible scenario. Other compatible scenario is associating 
the lower part to the filament over the PIL, under overarching loops (upper component 
more tenuous) with perpendicular fields to the first one. However, the magnetic 
skeleton of the flux rope-like filament observed by \cite{CK2009} can also 
reach low chromospheric layers \citep{yelles2012}. In this case, our model 
would suggest that the filament has strong fields in its lower layers and 
weak fields in the upper ones. In this case, 
using the magnetic topology inferred by \cite{yelles2012}, the ensuing vertical 
magnetic field strength gradient would be larger than 500 G/Mm. 

{ The key aspect of the solution found is the $90^\circ$ difference in azimuth 
between the top and bottom slabs. This extreme misalingment between 
the azimuths of both components is freely retrieved from the inversion since 
the profiles show no evidence of atomic polarization. 

After an exhaustive search, we were unable to find a solution where the two components 
have similar magnetic field inclinations but a difference in azimuth significantly smaller than 
$90^\circ$. We note that forcing more similar azimuths in the two
components ($\sim 30^\circ$) unavoidably leads to putting the  
magnetic field of the upper component roughly vertical ($\sim 150^\circ$), 
which is not favored by theoretical models. However, the quality of the fit
in this case is much worse, only compatible with the observations at the level of 10$^{-3}$.


}


In our opinion, two factors can contribute to the explanation of why 
sometimes Hanle-dominated signals are found while Zeeman-dominated 
profiles are found in other cases. First, the filament in \cite{CK2009} 
goes above penumbral regions of the active region, 
so that the field in the lower chromosphere turns out to be much higher. On the contrary, in 
our observations, the filament is above the granulation, so that the field and the absorption 
in the chromosphere are smaller. Second, there is a
dependence of the emergent profiles on the optical depth of the filament. 

Finally, we note that inferring the magnetic field properties of filaments under this model
will be more complicated because of the potential ambiguities. More work needs to be done
in order to improve the inversions, possibly using the full field-of-view to constrain
the model parameters.
{ However, we can safely say that a single-component inversion 
correctly retrieves: the total optical depth of the plasma in 
the filament (the difficult part is to separate it into two contributions) and the 
magnetic field of the dominant component (in our case, the lower component).
}

\begin{acknowledgments}
The authors are very grateful to Drs. C. Kuckein and R. Manso Sainz 
for very helpful discussions that improved the work and strengthened our conclusions. Financial support by the Spanish Ministry of Economy and Competitiveness 
through projects AYA2014-60476-P, AYA2014-60833-P and Consolider-Ingenio 2010 CSD2009-00038 
are gratefully acknowledged. C. J. D\'iaz Baso acknowledges Fundaci\'on La Caixa
for the financial support received in the form of a PhD contract. 
AAR also acknowledges financial support through the Ram\'on y Cajal fellowships. 
This research has made use of NASA's Astrophysics Data System Bibliographic Services.
\end{acknowledgments}

{
\begin{appendix}
\section{Aproximations in the radiative transfer problem}
\label{sec:app1}
In order to solve the radiative transfer problem we are assuming that the tensor components of the radiation field
$J^0_0$ (mean intensity) and $J^2_0$ (radiation anisotropy) that illuminate the top component are the same as those illuminating the
lower component. This is a necessary simplification in \Hazel\ to solve a linear problem in the statistical equilibrium equations. 
Otherwise, the full non-LTE problem of the second kind has to be solved, something that is well beyond our aims. However, in the following, we estimate the possible impact of the presence of the lower component on the illumination of the top component.

Concerning the radiation anisotropy of the upper component, $\left[J^2_0\right]_2$, the expression of this tensor component for the top 
slab in the plane-parallel case is given by:
\begin{equation}
\left[J_2^0\right]_{2} = \frac{1}{2 \sqrt{2}} \int_{-1}^1 \mathrm{d}\mu\,\int_{0}^\infty \mathrm{d}\nu (3\mu^2-1)\,\left[\phi_\nu\right]_{2}\,\left[I_\nu(\mu)\right]_{1},
\end{equation}
where $\left[\phi_\nu\right]_2$ is the line absorption profile in the upper component and $\left[I_\nu(\mu)\right]_1$ is the specific intensity at frequency $\nu$
and heliocentric angle $\mu$ that emerges from the lower component. Let us assume a Gaussian line
absorption profile with line width $\Delta \nu$ centered at frequency $\nu_0$. For simplicity, we
assume that the emergent intensity is an absorption Gaussian line with depression $d$ and that there is no 
differential Doppler shift between the two slabs (this case would generate atomic orientation). 
Also for simplicity, we assume that only the continuum has
center-to-limb variation ($d$ does not depend on $\mu$). Then, the ratio $r$ between 
the anisotropy of the upper slab illuminated with and without a spectral line is given by:
\begin{equation}
r=
1 - \frac{d}{\sqrt{2}}.
\end{equation}
Assuming values of $d \sim 0.25-0.4$ (this value is difficult to estimate because we only
observe the total absorption of the two components), the ratio yields values $r \sim 0.71-0.82$.
Therefore, the influence of the presence of an absorption spectral line on the lower component
seems to be negligible, specially given the difficulty in solving the full problem.

The second simplification in \Hazel\ is that the atomic orientation of the upper slab is zero. 
However, when the second layer is illuminated by circularly polarized light there is a non-zero
orientation of the radiation field $\left[J_1^0\right]_2$ which, in turn, produces atomic orientation in the top slab. 

In order to quantify $\left[J^1_0\right]_2$, 
\citet{marian2012} showed that it is possible to generate atomic
orientation, if there is a relative motion between the lower and upper component. The general
expression for the orientation of the radiation field is given by:
\begin{equation}
\left[J^1_0\right]_2 = \sqrt{\frac{3}{2}} \int_{-1}^1 \mathrm{d}\mu \int_{0}^\infty \mathrm{d}\nu \mu \left[\phi_\nu\right]_2 \left[V_\nu(\mu)\right]_1,
\end{equation}
where $V_\nu(\mu)$ is the emergent Stokes $V$ from the lower component. According to
\citet{marian2012}, if the relative velocity between the two components 
is of the order of 10-12 km s$^{-1}$, some atomic orientation can be generated in the upper
component. However, this would produce a visible symmetric Stokes $V$ profile, something
that we do not observe.
\end{appendix}
}
\clearpage

\bibliographystyle{apj}

\end{document}